\documentclass[epj-spec]{svjour}

\usepackage{graphics,epsfig,wrapfig,amssymb} 

\newcommand{\be}{\begin{equation}}
\newcommand{\ee}{\end{equation}} 
\newcommand{\bea}{\begin{eqnarray}}
\newcommand{\eea}{\end{eqnarray}}
\newcommand{\la}{\langle}
\newcommand{\ra}{\rangle}
\newcommand{\di}{{\rm d}}

\begin{document}

\graphicspath{{pretzel_eps/}}

\title{\boldmath Recent developments in nucleon spin structure\\ with focus on
 $h_{\rm 1L}^{\perp}$ and pretzelosity $h_{\rm 1T}^\perp$}

\subtitle{\it\normalsize Dedicated to 65 anniversary of Professor 
Klaus Goeke\thanks{The main part of the work was written during 
visiting by one of us (AE) Institut f\"ur Theoretische Physik II, 
Ruhr-Universit\"at Bochum. We thank Professor Klaus Goeke for his 
encouragement, collaboration and support, congratulate him 
with 65 anniversary and wish him good health for many years.}}

\author{H.~Avakian\inst{1} 
\and
\underline{A.\,V.~Efremov}\inst{2}\fnmsep\thanks{\email{efremov@theor.jinr.ru}} 
\and P.~Schweitzer{\inst{3,4}} 
\and A.~Metz\inst{5}
\and T.~Teckentrup\inst{3} 
}

\institute{Thomas Jefferson National Accelerator Facility,
Newport News, VA 23606, U.S.A. 
\and Joint Institute for Nuclear Research, Dubna, 141980 Russia 
\and Institut f\"ur Theoretische Physik II, Ruhr-Universit\"at
Bochum, Germany 
\and Department of Physics, University of Connecticut, 
Storrs, CT 06269, U.S.A.
\and Department of Physics, Barton Hall, Temple University, 
Philadelphia, PA 19122-6082, U.S.A. 
}

\titlerunning{Recent developments in nucleon spin structure}
\authorrunning{H.~Avakian, A. Efremov et al.}

\abstract{
The leading twist transverse momentum dependent parton 
distribution functions (TMDs) {$h_{\rm 1L}^{\perp}$} and 
$h_{\rm 1T}^\perp$, which is sometimes called ``pretzelosity,'' {are} 
studied. For $h_{\rm 1L}^{\perp}$ we consider {a 
``Wandzura--Wilczek-type'' {\sl approximation}, which follows from 
QCD equations of motion upon the neglect of pure twist-3 terms 
and allows to express it in terms of transversity}. On the basis 
of available data from HERMES we test the practical usefulness of 
this approximation and discuss how it can be further tested by 
future CLAS and COMPASS data.
We review the theoretical properties of pretzelosity and 
observe an interesting relation valid in a large class of 
relativistic models: The difference between helicity and 
transversity distributions, which is often said to be a 'measure 
of relativistic effects' in the nucleon, is nothing but the 
{transverse moment of} pretzelosity. 
We discuss preliminary deuteron target data from COMPASS on  
{the single spin asymmetry (SSA) in semi-inclusive deep 
inelastic scattering (SIDIS) related to pretzelosity,} 
and make predictions for future experiments at JLab, 
COMPASS and HERMES. }

\maketitle

\section{Introduction}
\label{intro} Processes like SIDIS, hadron production in $e^+e^-$ 
annihilations or the Drell--Yan process 
\cite{Cahn:1978se,Collins:1984kg,Sivers:1989cc,Efremov:1992pe,Collins:1992kk,Collins:1993kq,Kotzinian:1994dv,Mulders:1995dh,Boer:1997nt,Boer:1997mf,Boer:1997qn,Boer:1999mm,Brodsky:2002cx,Collins:2002kn,Belitsky:2002sm,Cherednikov:2007tw} 
factorize at leading twist 
\cite{Efremov:1980kz,Collins:1981uk,Ji:2004wu,Collins:2004nx} and 
allow to access information on transverse momentum dependent 
fragmentation functions and TMDs 
\cite{Collins:2003fm,Collins:2007ph}. The latter contain novel 
information on the nucleon structure. In order to be sensitive to 
``intrinsic'' transverse parton momenta it is necessary to 
measure adequate transverse momenta in the final state, for 
example, in SIDIS the transverse momenta of produced hadrons with 
respect to the virtual photon.

The eight leading-twist TMDs \cite{Boer:1997nt}, and further 
subleading-twist structures \cite{Goeke:2005hb,Bacchetta:2006tn} 
describe the structure of the nucleon in these reactions 
\be\label{Eq:TMD-pdfs}
    \underbrace{f_1^a,\, f_{\rm 1T}^{\perp a},\, g_{\rm 1L}^a,\,g_{\rm 1T}^a,\,
    h_{\rm 1T}^a,\, h_{\rm 1L}^{\perp a},\, h_{\rm 1T}^{\perp a},\,h_1^{\perp a},
    }_{\mbox{\footnotesize twist-2}}\,
    \underbrace{e^a,\,g_{\rm T}^a,\,h_{\rm L}^a,\,\dots }_{\mbox{\footnotesize twist-3}}
\ee
which are functions of $x$ and ${\bf p}_{\rm T}^2$. (The dots denote 
thirteen further twist-3 {TMDs}. The renormalization scale 
dependence is not indicated for brevity.) Integrating over 
transverse momenta one is left with six independent ``collinear'' 
{parton distribution functions (pdfs)} 
\cite{Ralston:1979ys,Jaffe:1991ra}
\be\label{Eq:coll-pdfs}
    \underbrace{f_1^a(x), \;\;\; g_1^a(x), \;\;\; h_1^a(x),}_{
    \mbox{\footnotesize twist-2}} \;\;\;
    \underbrace{  e^a(x), \;\;\; g_{\rm T}^a(x), \;\;\; h_{\rm L}^a(x).}_{
    \mbox{\footnotesize twist-3}}
\ee
where {we have}
$j(x) = \int\di^2{\bf p}_{\rm T} j(x,{\bf p}_{\rm T}^2)$ 
for $j=f_1^a,\,e^a,\,g_{\rm T},\,h_{\rm L}$ 
while $g_1^a(x) = \int\di^2{\bf p}_{\rm T} g_{\rm 1L}^a(x,{\bf p}_{\rm T}^2)$ and
$h_1^a(x) = \int\di^2{\bf p}_{\rm T} \{h_{\rm 1T}^a(x,{\bf p}_{\rm T}^2)+
{\bf p}_{\rm T}^2/(2M_N^2)
h_{\rm 1T}^{\perp a}(x,{\bf p}_{\rm T}^2)\}$.
 
The fragmentation of unpolarized hadrons is described in terms 
of two fragmentation functions, $D_1^a$ and $H_1^{\perp a}$, at 
leading-twist. In SIDIS (with polarized beams and/or targets, 
where necessary) it is possible to access information on the 
leading twist TMDs by measuring the angular distributions of 
produced hadrons. Some data on such processes are available 
\cite{Arneodo:1986cf,Airapetian:1999tv,Airapetian:2001eg,Airapetian:2002mf,Avakian:2003pk,Airapetian:2004tw,Alexakhin:2005iw,Diefenthaler:2005gx,Gregor:2005qv,Ageev:2006da,Avakian:2005ps,Airapetian:2005jc,Airapetian:2006rx,Abe:2005zx,Ogawa:2006bm,Martin:2007au,Diefenthaler:2007rj,Kotzinian:2007uv,Seidl:2008xc}.
The fragmentation functions and TMDs in SIDIS and other processes 
were subject to numerous studies in the literature 
\cite{Kotzinian:1995cz,DeSanctis:2000fh,Anselmino:2000mb,Efremov:2001cz,Efremov:2001ia,Ma:2002ns,Bacchetta:2002tk,Yuan:2003wk,Efremov:2003eq,D'Alesio:2004up,Anselmino:2005nn,Efremov:2004tp,Collins:2005ie,Collins:2005rq,Vogelsang:2005cs,Efremov:2006qm,Anselmino:2007fs,Arnold:2008ap,Anselmino:2008sg,Gamberg:2007gb,Kotzinian:2006dw,Avakian:2007mv,Avakian:2008dz,Metz:2008ib,Avakian:2007xa,Brodsky:2006hj,Burkardt:2007rv,Bacchetta:1999kz}. 
This is true especially for the prominent transversity 
distribution $h_1^a$ or the 'naively time-reversal-odd' functions 
like the Sivers function $f_{\rm 1T}^{\perp a}$, the Boer--Mulders 
function $h_1^{\perp a}$ and the Collins fragmentation function 
$H_1^{\perp a}$. {Among the so far less} considered functions are 
$h_{\rm 1L}^{\perp a}$ and the 'pretzelosity' distribution 
$h_{\rm 1T}^{\perp a}$.

The purpose of this lecture (based on the works 
\cite{Avakian:2007mv,Avakian:2008dz}) is fourfold. First, we 
discuss whether some of the unknown {TMDs} could be {{\sl 
approximated} in terms of (possibly better) known ones.} Second, 
we review what is known about $h_{\rm 1T}^{\perp a}$. Third, we 
mention the models these TMDs were calculated. 
Fourth, we present estimates for {SSAs} in which these 
functions enter, and discuss the prospects to measure these {SSAs}
in experiments at Jefferson Lab and COMPASS.

\begin{wrapfigure}[12]{R}{.5\textwidth}
\centering
\includegraphics[width=.5\textwidth]{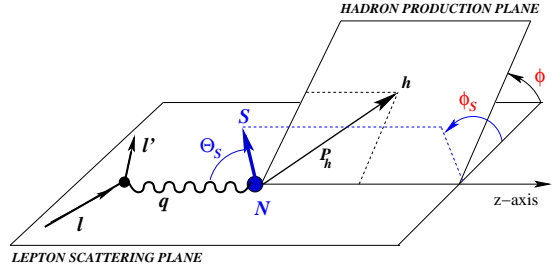}
\caption{\label{fig1-processes-kinematics}
Kinematics of SIDIS, $lN\to l^\prime h X$, 
and the definitions of azimuthal angles in the lab frame.}
\end{wrapfigure}

The process of SIDIS is sketched in 
Fig.~\ref{fig1-processes-kinematics}. We denote the momenta 
of the target, incoming and outgoing lepton by $P$, $l$ and $l'$ 
and introduce $s=(P+l)^2$, $q= l-l'$ 
with $Q^2= - q^2$. 
Then $y = \frac{Pq}{Pl}$, $x = \frac{Q^2}{2Pq}$, $z = 
\frac{PP_h}{Pq\;}$, and $\cos\theta_\gamma = 
1-\frac{2M_N^2x(1-y)}{s y}$ where $\theta_\gamma$ denotes the 
angle between target polarization vector and momentum ${\bf q}$ 
of the virtual photon $\gamma^\ast$, see 
Fig.~\ref{fig1-processes-kinematics}, and $M_N$ is the nucleon 
mass. The component of the momentum of the produced hadron 
transverse with respect to $\gamma^\ast$ is denoted by ${\bf 
P}_{\!h\perp}$ and $P_{h\perp}=|{\bf P}_{\!h\perp}|$.

The cross section differential in the azimuthal angle $\phi$ of 
the produced hadron has {schematically} the following general 
decomposition \cite{Kotzinian:1994dv,Diehl:2005pc} (the dots 
indicate power suppressed terms):
\bea\label{Eq:sigma-in-SIDIS}
\frac{\di\sigma}{\di\phi}
&=&                       F_{\rm UU}
\!+\cos(2\phi)    F_{\rm UU}^{\cos(2\phi)}
\!\!+S_{\rm L}\sin(2\phi) F_{\rm UL}^{\sin(2\phi)}
\!+\!\lambda\biggl[S_{\rm L}F_{\rm LL}
\!+S_{\rm T}\cos(\phi-\phi_S)F_{\rm LT}^{\cos(\phi-\phi_S)}\biggr]\nonumber\\
&&\hspace{-3mm}
+S_{\rm T}\biggl[\sin(\phi\!-\!\phi_S)F_{\rm UT}^{\sin(\phi\!-\!\phi_S)}
\!+\sin(\phi\!+\!\phi_S)F_{\rm UT}^{\sin(\phi\!+\!\phi_S)}
\!+\sin(3\phi\!-\!\phi_S)F_{\rm UT}^{\sin(3\phi\!-\!\phi_S)}\biggr]
\!+\dots\label{Eq:azim-distr-in-SIDIS}
\eea
In $F_{XY}^{\rm weight}$ the index $X={\rm U(L)}$ denotes the 
unpolarized (longitudinally polarized, helicity $\lambda$) beam. 
$Y={\rm U(L,T)}$ denotes the unpolarized target (longitudinally, 
transversely with respect to the virtual photon polarized 
target). The superscript reminds on the kind of angular 
distribution of the produced hadrons with no index indicating an 
isotropic $\phi$-distribution.

Each structure function arises from a different TMD. The 
chirally even $f$'s and $g$'s enter the observables in connection 
with the unpolarized fragmentation function $D_1^a$, the chirally 
odd $h$'s in connection with the chirally odd Collins 
fragmentation function~$H_1^{\perp a}$
\bea
\label{Eq:FUU}
F_{\rm UU} \propto\sum_ae_a^2\;f_{1 }^{a}\otimes\,D_1^{      a},\hspace{15mm}&&
F_{\rm LT}^{\cos( \phi-\phi_S)}\propto\sum_ae_a^2\;g_{\rm 1T}^{\perp a}\otimes\,D_1^{a}\,,\\
F_{\rm LL} \propto\sum_ae_a^2\;g_{1 }^{a}\otimes\,D_1^{a},\hspace{16mm}&&
F_{\rm UT}^{\sin( \phi-\phi_S)}\propto\sum_ae_a^2\;f_{\rm 1T}^{\perp a}\otimes\,D_1^{a}\,,\\
F_{\rm UU}^{\cos(2\phi)} \propto\sum_ae_a^2\;h_{1 }^{\perp a}\otimes\,H_1^{\perp a},\hspace{5mm}&&
F_{\rm UT}^{\sin( \phi+\phi_S)}\propto\sum_ae_a^2\;h_{1 }^{a}\otimes\,H_1^{\perp a}\,,\\
F_{\rm UL}^{\sin(2\phi)} \propto\sum_ae_a^2\;h_{\rm 1L}^{\perp a}\otimes\,H_1^{\perp a},\hspace{5mm}&&
F_{\rm UT}^{\sin(3\phi-\phi_S)}\propto\sum_ae_a^2\;h_{\rm 1T}^{\perp a}\otimes\,H_1^{\perp a}\,.
\label{AUTsin(3phi-phiS)}
\eea
More precisely, TMDs and fragmentation functions enter the 
respective {\sl tree-level} expressions in certain convolution 
integrals, indicated by $\otimes$ in 
(\ref{Eq:FUU}-\ref{AUTsin(3phi-phiS)}), which entangle transverse 
parton momenta from TMDs and fragmentation functions. (Going 
beyond tree-level description requires introduction of soft 
factors \cite{Ji:2004wu,Collins:2004nx} from which we refrain 
here.)
In general, such convolution integrals cannot be solved, unless 
one weighs the DIS counts with an adequate power of transverse 
hadron momentum
\cite{Boer:1997nt}.


\section{WW-type approximation for \boldmath $h_{\rm 1L}^{\perp}(x)$}
{It is clear that there cannot be any exact relations among TMDs,
because all TMDs are independent functions \cite{Goeke:2005hb}.
One may, however, find {\sl approximate} relations as follows.} 
From QCD equations of motion (eom), one obtains among 
others the following {\sl exact} relations \cite{Mulders:1995dh}
\be
    g_{\rm 1T}^{\perp(1)a}(x)\stackrel{\rm eom}{=}
    x\,g_{\rm T}^a(x)-x\,\tilde{g}_{\rm T}^a(x) \,,\label{Eq:eom-gT}\quad
    -2\,h_{\rm 1L}^{\perp(1)a}(x)\stackrel{\rm eom}{=}
    x\,h_{\rm L}^a(x)-x\,\tilde{h}_{\rm L}^a(x)
    \,,\label{Eq:eom-hL}
\ee
with 
$ h_{\rm 1L}^{\perp (1)a}(x) \equiv \int\di^2{\bf p}_{\rm T} 
    \frac{{\bf p}_{\rm T}^2}{{2M_N^2}}\;h_{\rm 1L}^{\perp a}(x,{\bf p}_{\rm T}^2)$,
$g_{\rm 1T}^{\perp(1)}$ analog,
and $\tilde{g}_{\rm T}^a(x)$, $\tilde{h}_{\rm L}^a(x)$ denoting pure 
twist-3 terms due to quark-gluon-quark 
correlations (and current quark mass terms). In the next step, we 
recall the relations among the collinear pdfs
(\ref{Eq:coll-pdfs}) 
\cite{Jaffe:1991ra,Wandzura:1977qf,Efremov:2002qh}
\be\label{Eq:WW-relation-gT}
   g_{\rm T}^a(x)=\int_x^1\frac{\di y}{y\;}\,g_1^a(y)+\tilde{g}_{\rm T}^{\prime a}(x)\,, 
   \label{Eq:WW-relation-hL}\quad
   h_{\rm L}^a(x)=2x\int_x^1\frac{\di y}{y^2\;}\,h_1^a(y)+\tilde{h}_{\rm L}^{\prime a}(x)
   \,,\ee
where $\tilde{g}_{\rm T}^{\prime a}(x)$, $\tilde{h}_{\rm L}^{\prime a}(x)$ 
denote pure twist-3 (and mass) terms 
\cite{Shuryak:1981pi,Jaffe:1989xx}, though different ones than 
in  (\ref{Eq:eom-hL}). Eqs.~(\ref{Eq:WW-relation-hL}) isolate 
``pure twist-3 terms'' in the ``twist-3'' pdfs $g_{\rm T}^a(x)$, 
$h_{\rm L}^a(x)$, because in (\ref{Eq:coll-pdfs}) the 
underlying ``working definition'' of twist \cite{Jaffe:1996zw} (a 
pdf is ``twist $t$'' if its contribution to the cross section is 
suppressed, in addition to kinematic factors, by $1/Q^{t-2}$ with 
$Q$ the hard scale) differs from the strict 
definition of twist (mass dimension of the operator minus its 
spin).

\begin{wrapfigure}[16]{R}{.32\textwidth}
\centering\vskip5mm
\includegraphics[width=.33\textwidth]{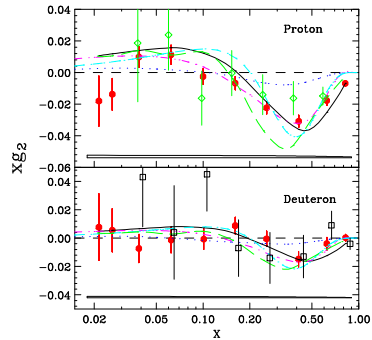}
\caption{\label{figg2WW}
WW-approximation for $g_2(x)\equiv \sum_ae_a^2\left(g_{\rm T}^a(x)- 
g_1^a(x)\right)$ in comparison with E155 data 
\cite{Anthony:2002hy}.}
\end{wrapfigure}

The remarkable observation is that $\tilde{g}_{\rm T}^{\prime a}(x)$ is 
consistent with zero within error bars 
\cite{Zheng:2004ce,Amarian:2003jy,Anthony:2002hy} and to a good 
accuracy 
\bea\label{Eq:WW-approx-gT}
    g_{\rm T}^a(x) \stackrel{\rm WW}{\approx}\int_x^1\frac{\di y}{y\;}\,g_1^a(y) 
    \;\;\;\mbox{(exp.\ observation)}
\eea
which is the ``Wandzura--Wilczek (WW) approximation'', whose
experimental status is demonstrated on Fig.~\ref{figg2WW}.

Lattice QCD \cite{Gockeler:2000ja,Gockeler:2005vw} and the 
instanton model of the QCD vacuum \cite{Balla:1997hf} support 
this observation. Interestingly the latter predicts also 
$\tilde{h}_{\rm L}^{\prime a}(x)$ to be small \cite{Dressler:1999hc}, 
such that
\be\label{Eq:WW-approx-hL}
    h_{\rm L}^a(x) \approx 2x\int_x^1\frac{\di y}{y^2}\,h_1^a(y) 
    \;\;\;\;\mbox{(prediction).}
\ee

On the basis of this positive experimental and theoretical 
experience one may suspect that the analog terms in the relations 
(\ref{Eq:eom-hL}) could also be negligible. If true one would 
have 
\bea
    g_{\rm 1T}^{\perp(1)a}(x)&\stackrel{\rm !?}{\approx}& \phantom{-}
    x\,\int_x^1\frac{\di y}{y\;}\,g_1^a(y) \;,
    \label{Eq:WW-approx-g1T}\\
        h_{\rm 1L}^{\perp(1)a}(x)&\stackrel{\rm !?}{\approx}& -
    x^2\!\int_x^1\frac{\di y}{y^2\;}\,h_1^a(y)\;,
    \label{Eq:WW-approx-h1L}\eea
that could be satisfied with an accuracy similar to 
(\ref{Eq:WW-approx-gT}). This remains to be tested in experiment.

An immediate application (or test) for the relations 
(\ref{Eq:WW-approx-g1T},~\ref{Eq:WW-approx-h1L}) is provided by 
the following single/double spin asymmetries (SSA/DSA) in SIDIS
\bea
     A_{\rm UL}^{\sin2\phi} &\propto&
    \sum\limits_ae_a^2\,h_{\rm 1L}^{\perp(1)a}\,H_1^{\perp a}\label{Eq:SSA-AUL2}\,,\\
        A_{\rm LT}^{\cos(\phi-\phi_S)} &\propto& 
    \sum\limits_ae_a^2\;g_{\rm 1T}^{\perp(1)a}\;D_1^a\,,\label{Eq:DSA-ALT}
\eea
where the Collins fragmentation function $H_1^{\perp a}$  
\cite{Efremov:1992pe,Collins:1992kk,Collins:1993kq} in  
(\ref{Eq:SSA-AUL2})  was determined very recently from SIDIS data 
\cite{Airapetian:2004tw,Alexakhin:2005iw,Diefenthaler:2005gx,Ageev:2006da} 
on the SSA 
\be\label{Eq:SSA-AUT}
        A_{\rm UT}^{\sin(\phi+\phi_S)}\propto\sum\limits_ae_a^2\,h_1^a\,H_1^{\perp a}
\ee
together with $e^+e^-$ annihilation data 
\cite{Abe:2005zx,Ogawa:2006bm} giving rise to a first but already 
consistent picture 
\cite{Vogelsang:2005cs,Efremov:2006qm,Anselmino:2007fs}. The     
$D_1^a$ in (\ref{Eq:DSA-ALT}) is the unpolarized fragmentation 
function which enters, of course, also the respective 
denominators in  (\ref{Eq:SSA-AUL2}-\ref{Eq:SSA-AUT}) that are 
proportional to $\sum_ae_a^2\,f_1^a\,D_1^a$.

Final HERMES 
\cite{Airapetian:1999tv,Airapetian:2001eg,Airapetian:2002mf} and 
preliminary CLAS \cite{Avakian:2005ps} data on 
(\ref{Eq:SSA-AUL2}) and preliminary COMPASS data 
\cite{Kotzinian:2007uv} on  (\ref{Eq:DSA-ALT}) are available, 
such that first tests of the WW-type approximations 
(\ref{Eq:WW-approx-g1T},~\ref{Eq:WW-approx-h1L}) are now or soon 
possible. A test of the approximation (\ref{Eq:WW-approx-g1T}) 
was suggested in \cite{Kotzinian:2006dw} along the lines of the 
study of the SSA (\ref{Eq:DSA-ALT}) discussed previously also in 
\cite{Kotzinian:1995cz}.

In this lecture we present a test of the approximation 
(\ref{Eq:WW-approx-h1L}). Under the assumption that this 
approximation {\sl works}, we shall see that it yields results 
for the SSA (\ref{Eq:SSA-AUL2}) compatible with data 
\cite{Airapetian:1999tv,Airapetian:2001eg,Airapetian:2002mf}. 
From another point of view our work provides a first independent 
cross check from SIDIS for the emerging picture of  $H_1^\perp$ 
\cite{Vogelsang:2005cs,Efremov:2006qm,Anselmino:2007fs}. The SSA 
(\ref{Eq:SSA-AUL2}) was recently studied in \cite{Gamberg:2007gb}.

Among the eight structure functions in SIDIS described in terms 
of twist-2 TMDs and fragmentation functions the SSAs 
(\ref{Eq:SSA-AUL2},~\ref{Eq:DSA-ALT}) are the only ones, for 
which WW-type approximations could be of use. Exact eom-relations 
exist for all eight twist-2 pdfs in (\ref{Eq:TMD-pdfs}). But the 
relations (\ref{Eq:eom-hL}) are special in that they connect the 
respective TMD pdfs, namely $g_{\rm 1T}^\perp$ and $h_{\rm 1L}^\perp$, to 
``collinear'' twist-3 pdfs, namely $g_{\rm T}$ and $h_{\rm L}$. Those in turn 
are related to twist-2 pdfs, $g_1$ and $h_1$, by means of 
(experimentally established or theoretically predicted) 
WW-approximations (\ref{Eq:WW-approx-gT},~\ref{Eq:WW-approx-hL}).

Experiments may or may not confirm that the  WW-type 
approximations (\ref{Eq:WW-approx-g1T},~\ref{Eq:WW-approx-h1L}) 
work.

What would it mean if 
(\ref{Eq:WW-approx-g1T},~\ref{Eq:WW-approx-h1L}) were found to be 
satisfied to within a very good accuracy? First, that would be of 
practical use for understanding and interpreting the first data 
\cite{Airapetian:1999tv,Airapetian:2001eg,Airapetian:2002mf,Avakian:2003pk,Airapetian:2004tw,Alexakhin:2005iw,Diefenthaler:2005gx,Gregor:2005qv,Ageev:2006da,Avakian:2005ps,Airapetian:2005jc,Airapetian:2006rx,Abe:2005zx,Ogawa:2006bm,Martin:2007au,Diefenthaler:2007rj,Kotzinian:2007uv}. 
Second, it would call for theoretical explanations why pure 
twist-3 terms should be small. (Only for the smallness of the 
``collinear'' pure twist-3 terms in 
(\ref{Eq:WW-approx-gT},~\ref{Eq:WW-approx-hL}) lattice QCD 
\cite{Gockeler:2000ja,Gockeler:2005vw} and/or instanton vacuum 
\cite{Balla:1997hf,Dressler:1999hc} provide explanations.)

What would it mean if 
(\ref{Eq:WW-approx-g1T},~\ref{Eq:WW-approx-h1L}) were found to 
work poorly? This scenario would be equally interesting. In fact, 
all TMDs in (\ref{Eq:TMD-pdfs}) are independent structures, 
and any of them contains different type of information on the 
internal structure of the nucleon. The measurement of the 
complete set of all eighteen structure functions available in 
SIDIS \cite{Kotzinian:1994dv} is therefore indispensable for our 
aim to learn more about the nucleon structure.

One type of information accessible in this way concerns effects 
related to the orbital motion of quarks, and in particular 
correlations of spin and transverse momentum of quarks which are 
dominated by valence quarks and hence play a more important role 
at large $x$. E.g.~it was shown that spin-orbit correlations may 
lead to significant contributions to partonic momentum and 
helicity distributions \cite{Avakian:2007xa} in the large-$x$ limit. 
Spin-orbit correlations are presumably of similar importance for 
transversity, and crucial for $h_{\rm 1L}^\perp$, which 
is a measure for the correlation of the transverse spin and the 
transverse momentum of quarks in a longitudinally polarized 
nucleon.


\section{\boldmath Properties and experimental check of WW-type 
approximation for \boldmath $h_{\rm 1L}^\perp$} 
\label{Sec-2:h1Lperp-in-WW-approx}

\begin{figure}[h!]
\vspace{-0.7cm}\centering
        \includegraphics[width=0.3\textwidth]{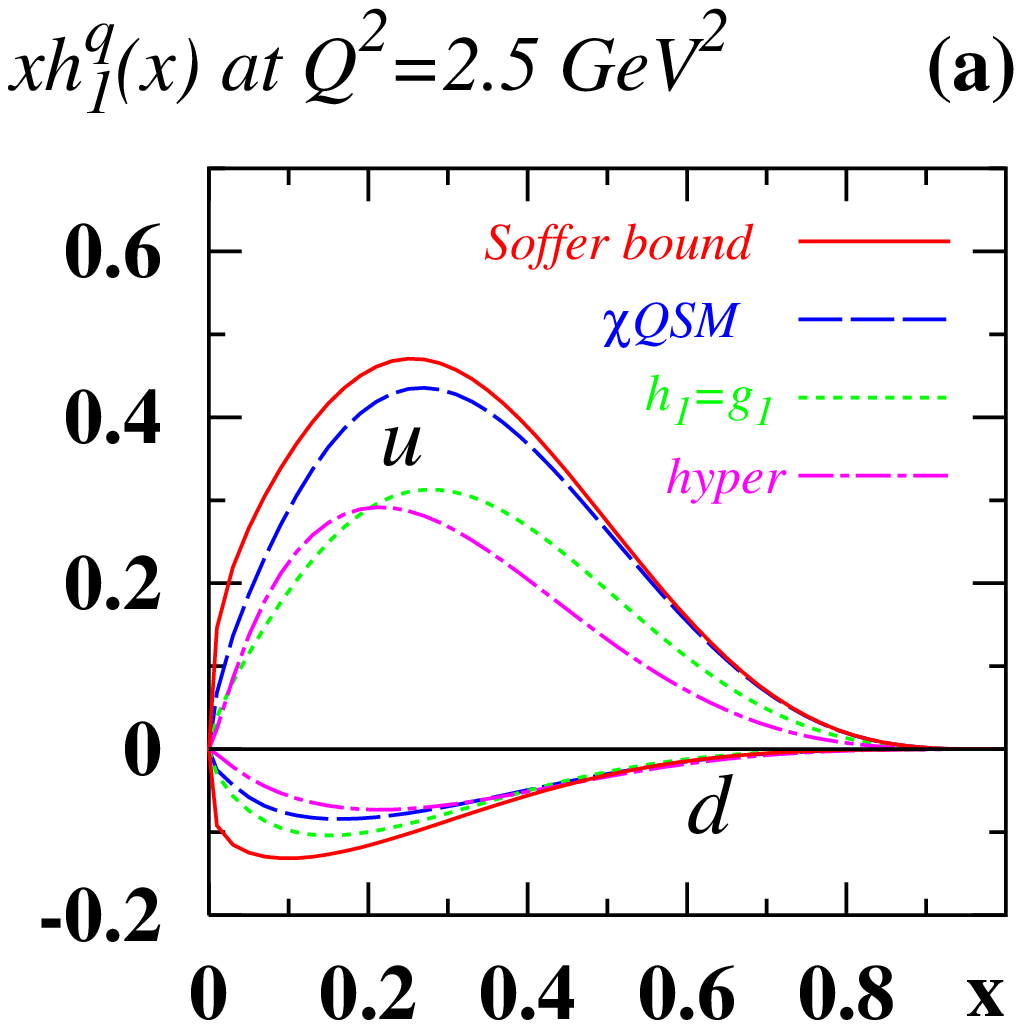}
        \includegraphics[width=0.3\textwidth]{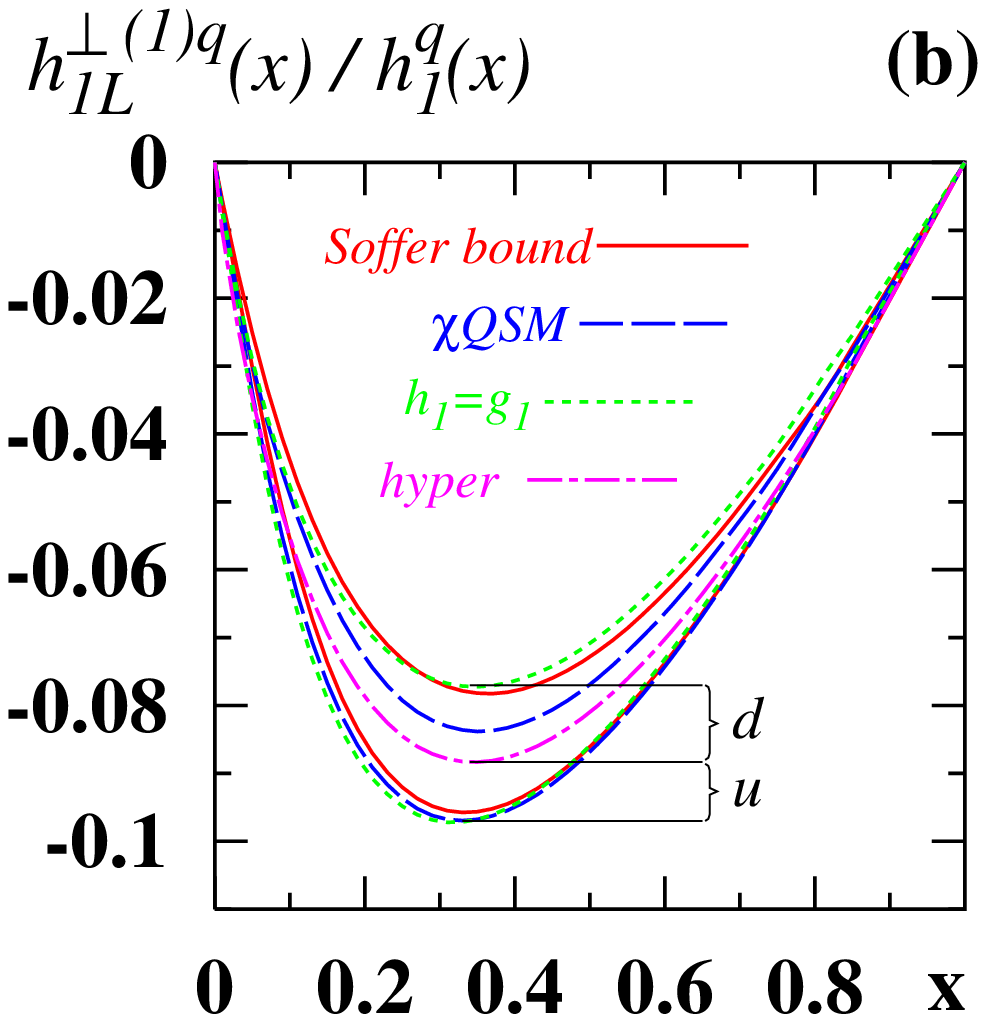}
        \includegraphics[width=0.3\textwidth]{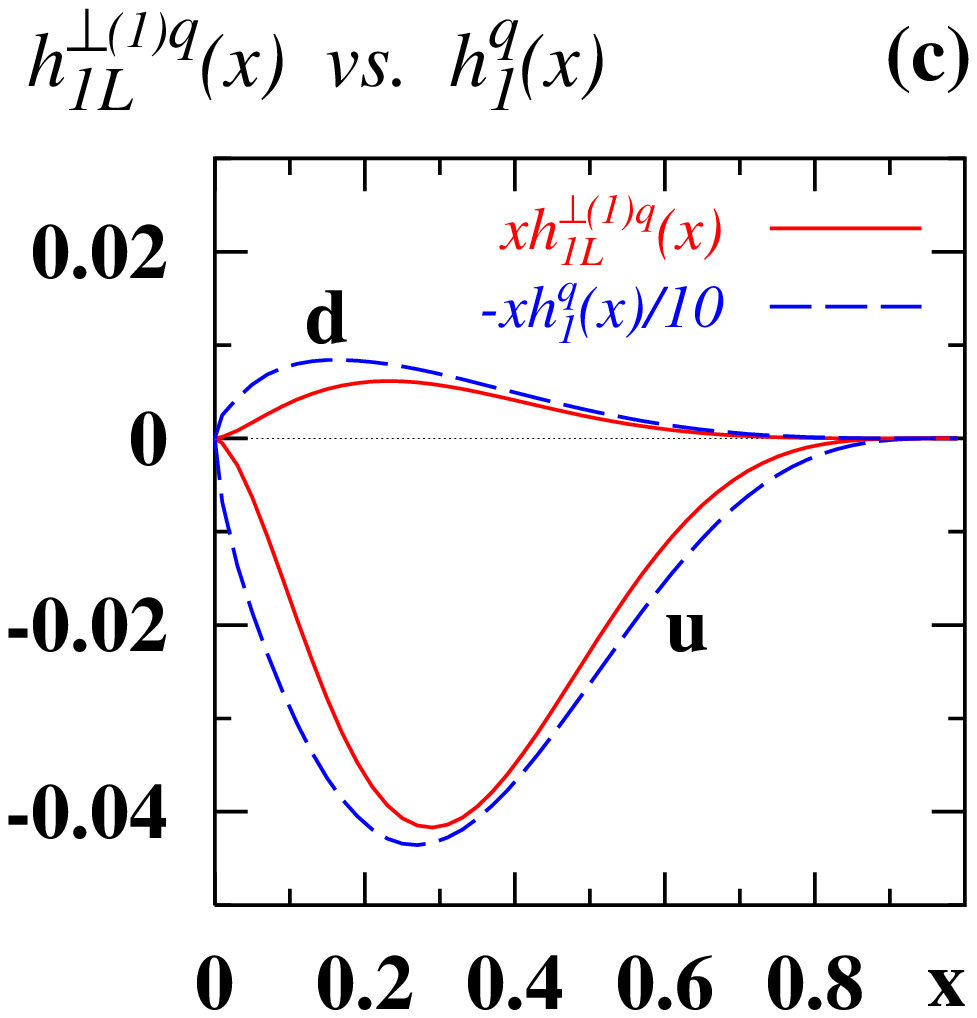}
\caption{\label{Fig01:h1Lperp-x}
    (a) Transversity, $xh_1^a(x)$, vs.\ $x$, from various models.
    (b) The ratio $h_{\rm 1L}^{\perp(1)a}(x)/h_1^a(x)$ vs.\ $x$ in 
    various models, with $h_{\rm 1L}^\perp$ estimated by means of the
    WW-type approximation (\ref{Eq:WW-approx-h1L}).
    (c) $xh_{\rm 1L}^{\perp(1)a}(x)$ vs.\ $x$ from the WW-type 
    approximation (\ref{Eq:WW-approx-h1L}) and $h_1^a(x)$ from $\chi$QSM 
    \cite{Schweitzer:2001sr}, in comparison with $(-\,\frac{1}{10})xh_1^a(x)$
    from that model. All results here refer to a scale of $2.5\,{\rm GeV}^2$. }
\end{figure}

In order to model $h_{\rm 1L}^{\perp(1)a}(x)$ by means of the WW-type 
approximation (\ref{Eq:WW-approx-h1L}) one inevitably has to use, 
in addition, models for the transversity pdf. 
Fig.~\ref{Fig01:h1Lperp-x}a shows four different models: 
saturation of the Soffer bound \cite{Soffer:1994ww} at the low 
initial scale of the leading order parameterizations 
\cite{Gluck:1998xa} (choosing $h_1^u>0$ and $h_1^d<0$), the 
chiral quark soliton model ($\chi$QSM) \cite{Schweitzer:2001sr}, 
the non-relativistic model assumption $h_1^a(x)=g_1^a(x)$ at the 
low scale of the parameterization \cite{Gluck:1998xa}, and the 
hypercentral model \cite{Pasquini:2005dk}. All curves in 
Fig.~\ref{Fig01:h1Lperp-x} are leading-order evolved to 
$2.5\,{\rm GeV}^2$ which is a relevant scale in experiment, see 
below.

These (and many other \cite{Barone:2001sp,Efremov:2004tz}) models 
agree on that $h_1^u(x)>0$ and $h_1^d(x)<0$ with $|h_1^d(x)| < 
h_1^u(x)$, though the predictions differ concerning the 
magnitudes, see Fig.~\ref{Fig01:h1Lperp-x}a. Models in which 
antiquark distribution functions can be computed, e.g.\  
\cite{Schweitzer:2001sr}, predict that the transversity antiquark 
pdfs are far smaller than the quark ones.

Let us establish first a robust feature of the relation 
(\ref{Eq:WW-approx-h1L}), namely the ratio 
$R=h_{\rm 1L}^{\perp(1)a}(x)/h_1^a(x)$ exhibits little dependence on 
the transversity model, see Fig.~\ref{Fig01:h1Lperp-x}b. A 
``universal'' behaviour of this ratio at large $x$ is not 
surprising. By inspecting (\ref{Eq:WW-approx-h1L}) one can prove: 
\begin{itemize}
\item for $x\to 1$, it behaves like $R\approx (1-x)$,
\item for $x\to 0$, it also vanishes: $R\to 0$,
\item a  common feature: $|R|\lesssim 0.1$.
\end{itemize}
In the following we will use the $\chi$QSM 
\cite{Diakonov:1987ty,Christov:1995vm}, see 
Fig.~\ref{Fig01:h1Lperp-x}c, which has several advantages: it 
describes the twist-2 pdfs $f_1^a(x)$ and $g_1^a(x)$ within 
(10--30)$\,\%$ accuracy \cite{Diakonov:1996sr}, it is derived from 
the instanton vacuum model \cite{Diakonov:1983hh,Diakonov:1995qy} 
which predicts the ``collinear WW-type approximation" 
(\ref{Eq:WW-approx-hL}) to work well \cite{Dressler:1999hc}, and 
below we will use information on the Collins effect from the analysis 
\cite{Efremov:2006qm} where this model was used. This helps to 
minimize the model-dependence in our study. But we shall see that 
our conclusions do not depend on the choice of model. 

{The SSA as defined in the HERMES experiment is given by}
\be\label{Eq:AUL2-theory}
    A_{\rm UL}^{\sin2\phi} = 
    \frac{\sum_i\sin(2\phi_i)(N_i^{\leftrightarrows}-N_i^\rightrightarrows)}
    {\sum_i\frac12(N_i^{\leftrightarrows}+N_i^\rightrightarrows)}= 
    \frac{\int\di y\,[\cos\theta_\gamma(1-y)/Q^4] F_{\rm UL}^{\sin2\phi}}
             {\int\di y\,[      (1-y+\frac12y^2)/Q^4] F_{\rm UU}}
\ee
where $N_i^{\leftrightarrows}$ ($N_i^\rightrightarrows$) denotes 
the number of events $i$ with target polarization antiparallel 
(parallel) to the beam. Had the events in the numerator of 
(\ref{Eq:AUL2-theory}) been weighted by $P_{h\perp}^2/(M_N m_h)$ 
in addition to $\sin(2\phi)$, the convolution integral could be 
solved in a model independent way with the result given in terms 
of the transverse moment of $h_{\rm 1L}^\perp$ and an analog moment 
for $H_1^\perp$ \cite{Boer:1997nt}. Including such an additional 
weight makes data analysis more difficult due to acceptance 
effects. Omitting it, however, forces one to resort to models.

Here we shall assume the distributions of transverse parton 
momenta to be Gaussian. 
Such an Ansatz satisfactorily describes data on many hard 
reactions \cite{D'Alesio:2004up}, provided the transverse momenta 
are much smaller than the hard scale of the process, i.e.\  $\la 
P_{h\perp}\ra\ll \la Q\ra$ which is the case at HERMES. In fact, 
the $z$-dependence of $\la P_{h\perp}\ra$ at HERMES 
\cite{Airapetian:2002mf} is well described in the Gauss Ansatz 
\cite{Collins:2005ie}.
Of course, one has to keep in mind that it 
is a crude approximation, and it is not clear whether it works 
also for polarized distribution and fragmentation functions.
Moreover, since also unintegrated versions of (\ref{Eq:eom-hL}) hold, 
this Ansatz cannot be equally valid for all TMDs.

\begin{figure}[h]
\centering\vskip-5mm
\includegraphics[width=.27\textwidth]{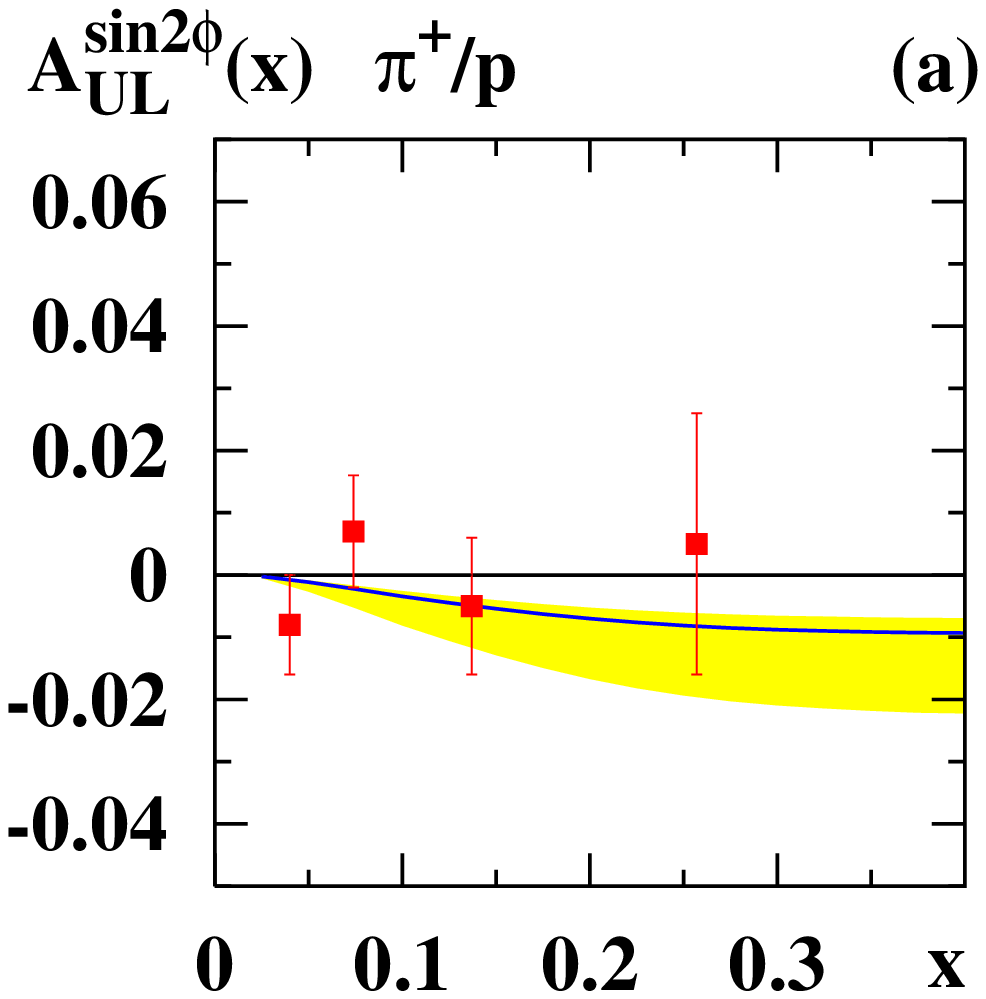}\hskip-5mm
\includegraphics[width=.27\textwidth]{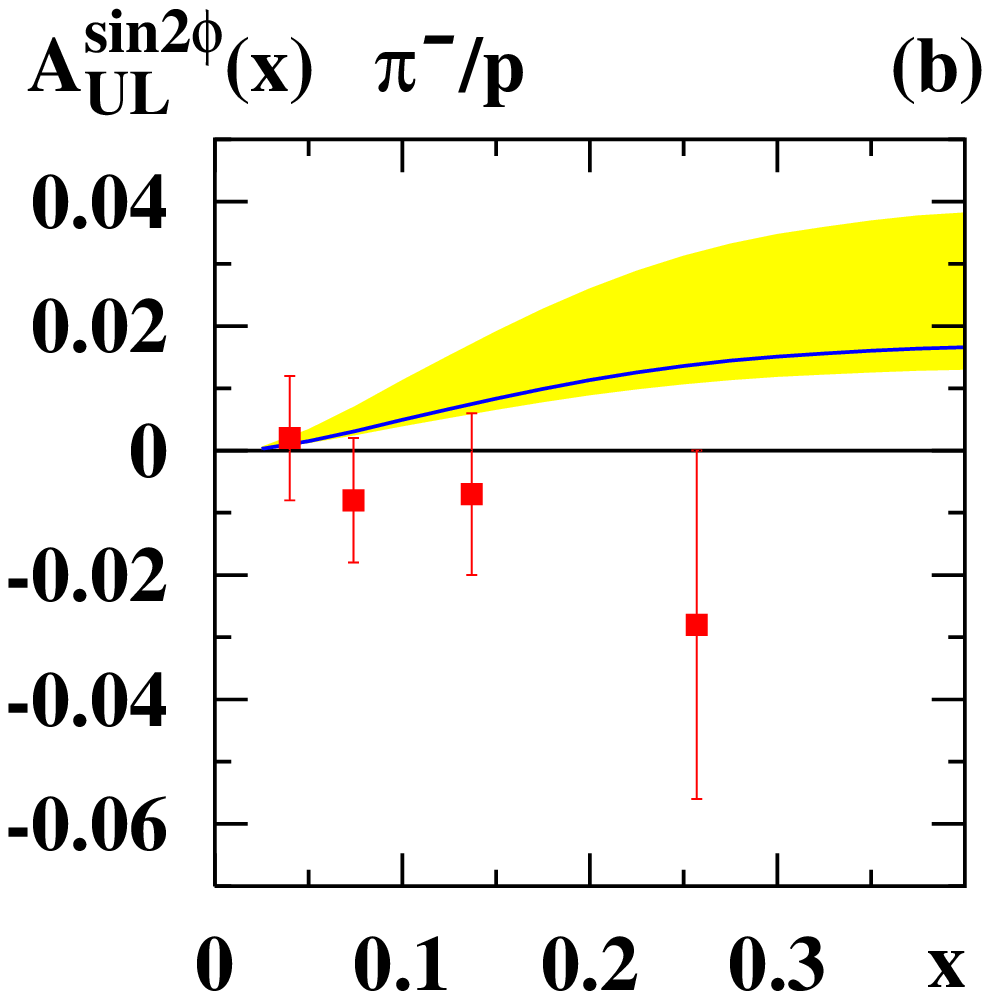}\hskip-5mm
\includegraphics[width=.27\textwidth]{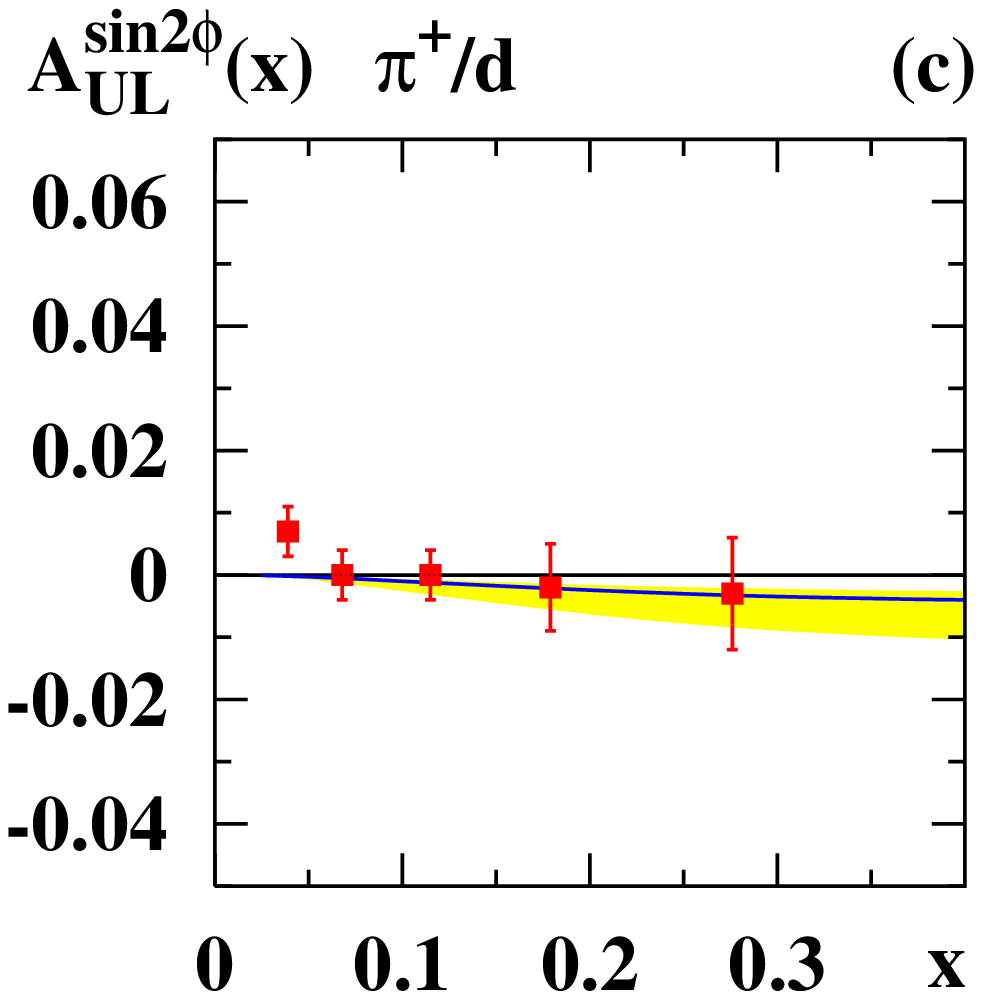}\hskip-5mm
\includegraphics[width=.27\textwidth]{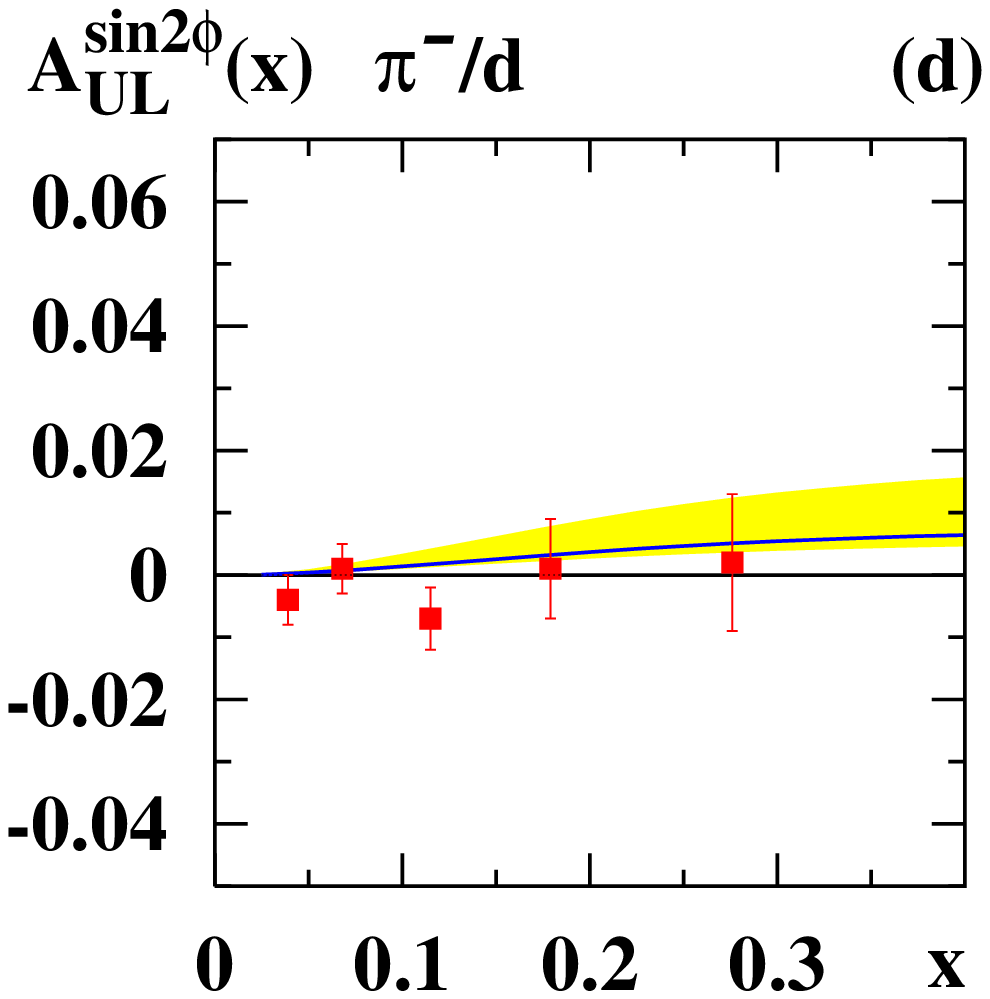}
\caption{\label{Fig03:AUL2-x}
Longitudinal target SSA $A_{\rm UL}^{\sin2\phi}$ as function of~$x$. 
The proton (a, b) and deuterium (c,d) target data  are from 
HERMES \cite{Airapetian:1999tv,Airapetian:2002mf}. The 
theoretical curves are obtained using information on $H_1^\perp$ 
\cite{Efremov:2006qm,Anselmino:2007fs}, predictions from the 
instanton vacuum model and chiral quark soliton model for $h_{\rm L}^a$ 
and $h_1^a$ \cite{Dressler:1999hc,Schweitzer:2001sr}, and --- 
this is crucial in our context --- assuming the validity of the 
WW-type approximation (\ref{Eq:WW-approx-h1L}). The shaded error 
bands are due to the uncertainties in $H_1^\perp$, see 
\cite{Avakian:2007mv} for details.}
\end{figure}

What is convenient for our purposes is that the Gauss Ansatz 
allows to solve the convolution integral. Using the WW-type 
approximation (\ref{Eq:WW-approx-h1L}) with $h_1^a(x)$ from the 
$\chi$QSM \cite{Schweitzer:2001sr}, and information on the Collins 
effect from 
\cite{Vogelsang:2005cs,Efremov:2006qm,Anselmino:2007fs}, we 
obtain the results shown in Fig.~\ref{Fig03:AUL2-x}. The error 
bands of the theoretical curves reflect the present uncertainties 
in the quantitative understanding of the Collins effect, see 
\cite{Avakian:2007mv} for explicit expressions and details.

%

 Our results shown in Fig.~\ref{Fig03:AUL2-x} for pion 
production from proton and deuteron targets are consistent with 
the HERMES data 
\cite{Airapetian:1999tv,Airapetian:2001eg,Airapetian:2002mf}, and 
do not exclude that (\ref{Eq:WW-approx-h1L}) is a useful 
approximation.

{Further insights are expected from the CLAS experiment at 
Jefferson Lab, which promises higher statistics due to two orders 
of magnitude higher luminosity, and provides access to much 
larger $x$ and larger $z$ than HERMES and COMPASS. Large $\la 
z\ra$} may also enhance the SSA due to Collins function 
$H_1^{\perp(1/2)a}(z) \propto zD_1^a(z)$, as  observed in 
\cite{Efremov:2006qm}. This makes CLAS an ideal experiment for 
studies of this SSA in particular and spin-orbit correlations in 
general. Comparison of the various data sets will also allow to 
draw valuable conclusions on the energy dependence of the 
process, possible power-corrections, etc.

The preliminary data from CLAS \cite{Avakian:2005ps} have shown 
non-zero SSAs for charged pions, and a compatible with zero 
within error bars result for $\pi^0$. Within our approach it is 
possible to understand the results for $\pi^+$ and $\pi^0$, 
however, we obtain for $\pi^-$ an opposite sign compared to the 
data. In view of this observation, it is worth to look again on 
Fig.~\ref{Fig03:AUL2-x}b which shows HERMES data on the 
$\pi^-$-SSA. Does Fig.~\ref{Fig03:AUL2-x}b hint at an 
incompatibility? Charged pions and in particular the $\pi^-$ may 
have significant higher twist contributions, in particular from 
exclusive vector mesons and semi-exclusive pion production at 
large $z$. 

New data expected from CLAS with $E_{\rm beam}=6\,{\rm GeV}$ 
\cite{avak-clas6}, will increase the existing statistics by about 
an order of magnitude and more importantly provide comparable to 
$\pi^+$ sample of $\pi^0$ events. Neutral pion sample is not 
expected to have any significant contribution from exclusive 
vector mesons, neither it is expected to have significant higher 
twist corrections due to semi-exclusive production of pions with 
large $z$ \cite{Afanasev:1996mj}, where the separation between 
target and current fragmentation is more pronounced. The JLab 
upgrade to 12\,GeV \cite{avak-clas12} will allow to run at an 
order of magnitude higher luminosity than current CLAS, providing 
a~comprehensive set of single and double spin asymmetries 
covering a wide range in $x$ and $z$. That will allow detailed 
studies of kinematic dependence of target SSA and clarify the 
situation.

COMPASS has taken data with a longitudinally polarized deuterium 
target which are being analyzed, and soon also a proton target 
will be used. The $160\,{\rm GeV}$ muon beam allows to extend the 
measurements of SSAs to small $x$. With the cut $Q^2> 1\,{\rm 
GeV}^2$ the average $\la Q^2\ra $ at COMPASS is comparable to 
that at HERMES. Therefore Fig.~\ref{Fig03:AUL2-x} shows roughly 
our predictions for COMPASS.
One may expect $A_{\rm UL}^{\sin2\phi}$ to be substantially smaller 
than $A_{\rm UT}^{\sin(\phi+\phi_S)}$, especially at small $x$. It 
will be interesting to see whether these predictions will be 
confirmed by COMPASS.


\section{\boldmath What do we know about $h_{\rm 1T}^{\perp}$?}

This TMD is defined as the coefficient of the correlation 
$\frac{1}{M^2}[(\vec S_{\rm T}\vec p_{\rm T})(\vec p_{\rm T}\vec s_{\rm T})-
\frac{1}{2}\vec p_{\rm T}^2(\vec S_{\rm T}\vec s_{\rm T})]$, where 
$\vec S_{\rm T}$ and $\vec s_{\rm T}$ denote the spin vectors of 
proton and quark, which are transverse with respect 
to the virtual photon momentum.
Let us summarize briefly, what we know about pretzelosity.
\begin{enumerate}
\renewcommand{\labelenumi}{\sl (\arabic{enumi})$\,$}
\item\label{point-h1Tperp-gluon}   
It is chirally odd and so there is no gluon analog of 
pretzelosity\footnote{Actually, in the decomposition of the gluon 
analog of the above correlator structure appears in 
Ref.~\cite{Meissner:2007rx} has been called $h_{\rm 1T}^{\perp g}$. 
(In Ref.~\cite{Mulders:2000sh} it was given a different name.) 
This gluon TMD, however, has different properties compared to our 
quark pretzelosity. For example, the $h_{\rm 1T}^{\perp g}$ of 
\cite{Meissner:2007rx} is `odd' under time-reversal while 
$h_{\rm 1T}^{\perp a}$ with $a=q,\;\bar q$ is `even'.}.
\item   It has a probabilistic interpretation \cite{Mulders:1995dh}.
\item   At large-$x$ it is predicted to be suppressed by $(1-x)^2$ compared to
$f_1^a(x)$ \cite{Avakian:2007xa,Brodsky:2006hj,Burkardt:2007rv}.
\item\label{point-small-x}
    It is expected to be suppressed also at small $x$ compared to $f_1^a$.
\item\label{point-positivity}
    It must satisfy the positivity condition \cite{Bacchetta:1999kz}:
\be\label{Eq:positivity}
    \biggl|h_{\rm 1T}^{\perp(1)a}(x,\vec{p}_{\rm T}^2)\biggr|
    \le\frac12\biggl(f_1^a(x,\vec{p}_{\rm T}^{2})-g_1^a(x,\vec{p}_{\rm T}^{2})\biggr)
\mbox{ ~or~ }|h_{\rm 1T}^{\perp(1)a}(x)|
    \le\frac12\biggl(f_1^a(x)-g_1^a(x)\biggr)\;.
    \label{Eq:positivity-integrated}
\ee
With the Soffer inequality
$|h_1^a(x)|\le\frac12(f_1^a + g_1^a)(x)$ \cite{Soffer:1994ww},
we obtain the remarkable bound:
    \be\label{Eq:positivity-with-h1}
    | h_{\rm 1T}^{\perp(1)a}(x)|+ |h_1^a(x)| \le f_1^a(x)\;.
    \ee
\item\label{point-large-Nc}
In the limit of a large number of colors $N_c$ it was shown that 
for $xN_c\approx {\cal O}(1)$ and $p_{\rm T}\approx {\cal O}(1)$ (the same 
pattern holds also for antiquarks) \cite{Pobylitsa:2003ty}:
\be\label{Eq:large-Nc} 
\frac{h_1^{\perp u}+h_1^{\perp d}}{h_1^{\perp u}-h_1^{\perp d}}
\approx {\cal O}(1/N_c)
\ee 
\item   It requires the presence of nucleon wave-function components with
   two units orbital momentum difference, e.g.\  $s$-$d$ interference
    or quadratic in $p$-wave component \cite{Burkardt:2007rv}.
\item   In some sense it `measures' the deviation of the `nucleon shape'
   from a sphere \cite{Miller:2007ae} (that is why it was called pretzelosity).
\item   In simple (spectator-type) models, it has been related to chirally
    odd generalized parton distributions \cite{Meissner:2007rx}.
\item\label{point-measure-of-relativity}   
It was observed in the bag model \cite{Avakian:2008dz} that
\be\label{Eq:measure-of-relativity}  
\fbox{$h_{\rm 1T}^{\perp(1)a}(x) = g_1^a(x) - h_1^a(x)$}
\ee
and since `the difference of $g_1^a(x)$ and $h_1^a(x)$ is a 
measure for relativistic effects in nucleon' \cite{Jaffe:1991ra} 
one may view the transverse moment of pretzelosity as a measure 
for relativistic effects.
\end{enumerate}

An important question is: How general the relation 
(\ref{Eq:measure-of-relativity})? It is clear that this relation 
cannot be strictly valid in QCD, where all TMDs are independent. 
However, could it nevertheless allow to gain a reasonable 
estimate for $h_{\rm 1T}^{\perp(1)}(x)$ in terms of transversity and 
helicity? Until clarified by experiment, we can address this 
question only in models.

Interestingly, the relation (\ref{Eq:measure-of-relativity}) does 
not only hold in the bag model \cite{Avakian:2008dz}, but is 
found \cite{Avakian:2008dz} to be satisfied also in the spectator 
model \cite{Jakob:1997wg}. In fact, it was conjectured 
\cite{Avakian:2008dz} that (\ref{Eq:measure-of-relativity}) could 
hold in a large class of {\sl relativistic quark models}, which 
was subsequently confirmed in the constituent quark model 
\cite{Pasquini:2008ax} and the relativistic model of the proton 
\cite{Efremov:2008mp} (but not in a variant of the spectator 
model \cite{Bacchetta:2008af}). These are all quark models. The 
limitations of (\ref{Eq:measure-of-relativity}) are nicely 
illustrated in the `quark target model' where in addition to 
quarks there are also gluons, and 
(\ref{Eq:measure-of-relativity}) is not satisfied 
\cite{Meissner:2007rx}. Thus, the explicit inclusion of gluon 
degrees of freedom spoils this relation, i.e.\ it can be valid in 
`no-gluon-models' only.

It would be interesting and instructive to know the necessary and 
sufficient conditions for the relation 
(\ref{Eq:measure-of-relativity}) to hold in a (quark) model.

\section{Preliminary COMPASS data \& prospects at JLab}

In SIDIS pretzelosity gives rise (in combination with the
Collins fragmentation function \cite{Collins:1992kk,Efremov:1992pe}) 
to the SSA 
\be\label{Eq:AUT-Gauss}
    A_{\rm UT}^{\sin(3\phi-\phi_S)} =
    \frac{C_{\rm Gauss}\sum_a e_a^2 x \,h_{\rm 1T}^{\perp(1) a}(x)\,
    H_1^{\perp (1/2)a}(z)} {\sum_ae^a\,xf_1^a(x) \,D_1^a(z)}
    \ee
where we assumed the Gauss Ansatz, whose parameters are contained 
in the factor $C_{\rm Gauss}$. This factor is a slowly varying 
function of the Gauss model parameters and can be well 
approximated for practical purposes by its maximum $C_{\rm Gauss} 
\le C_{\rm max} = 3/(2\sqrt{2})$, see for further details 
\cite{Avakian:2008dz}. 

At COMPASS the $\sin(3\phi-\phi_S)$ and other SSAs were measured 
on a deuteron target \cite{Kotzinian:2007uv}. By saturating the 
positivity bound for the transverse moment of pretzelosity 
(point~(5) in Sec.~4) we estimated \cite{Avakian:2008dz} the 
maximum effect for the SSA. We used the parameterizations 
\cite{Gluck:1998xa,Kretzer:2000yf}, and for $H_1^\perp$ the 
information from \cite{Efremov:2006qm,Anselmino:2007fs}.

%
\begin{figure}[h]
\vskip-8mm\centering
\begin{tabular}{ccc}
\includegraphics[width=50mm]{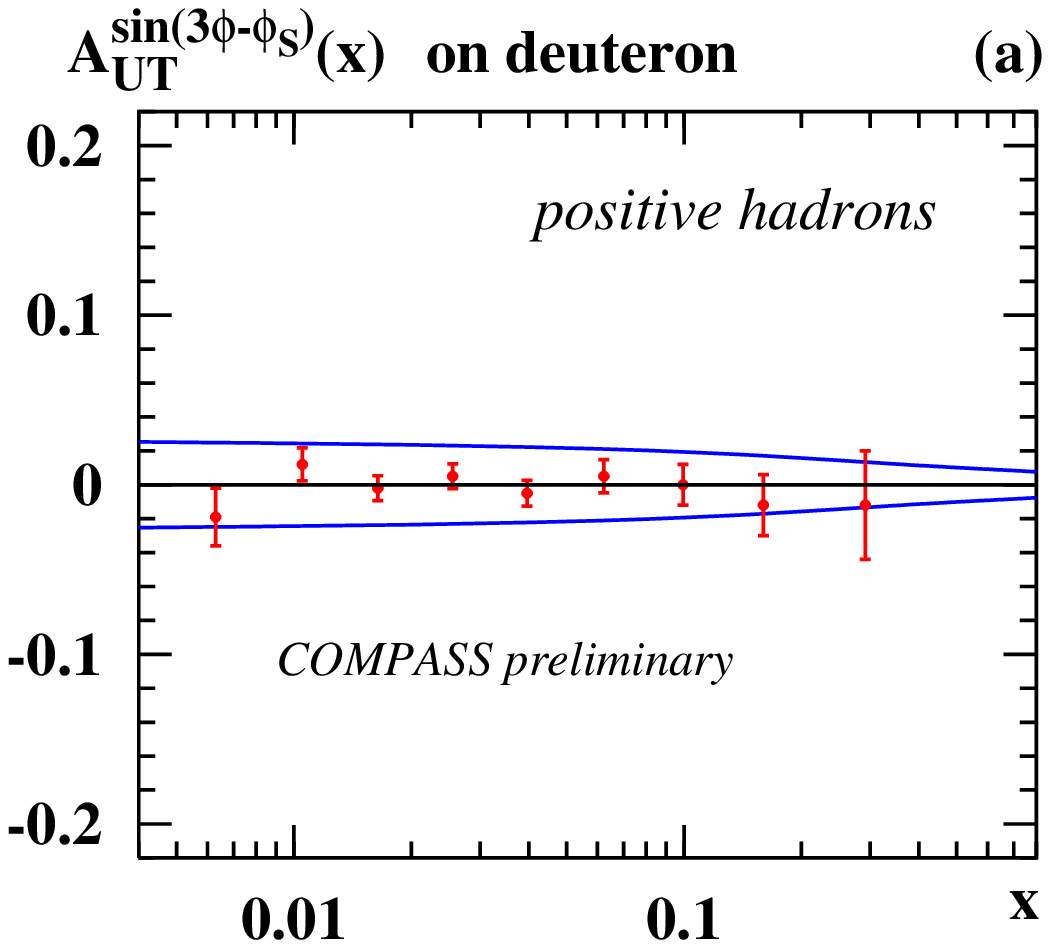} &
\hspace{-8mm}
\includegraphics[width=50mm]{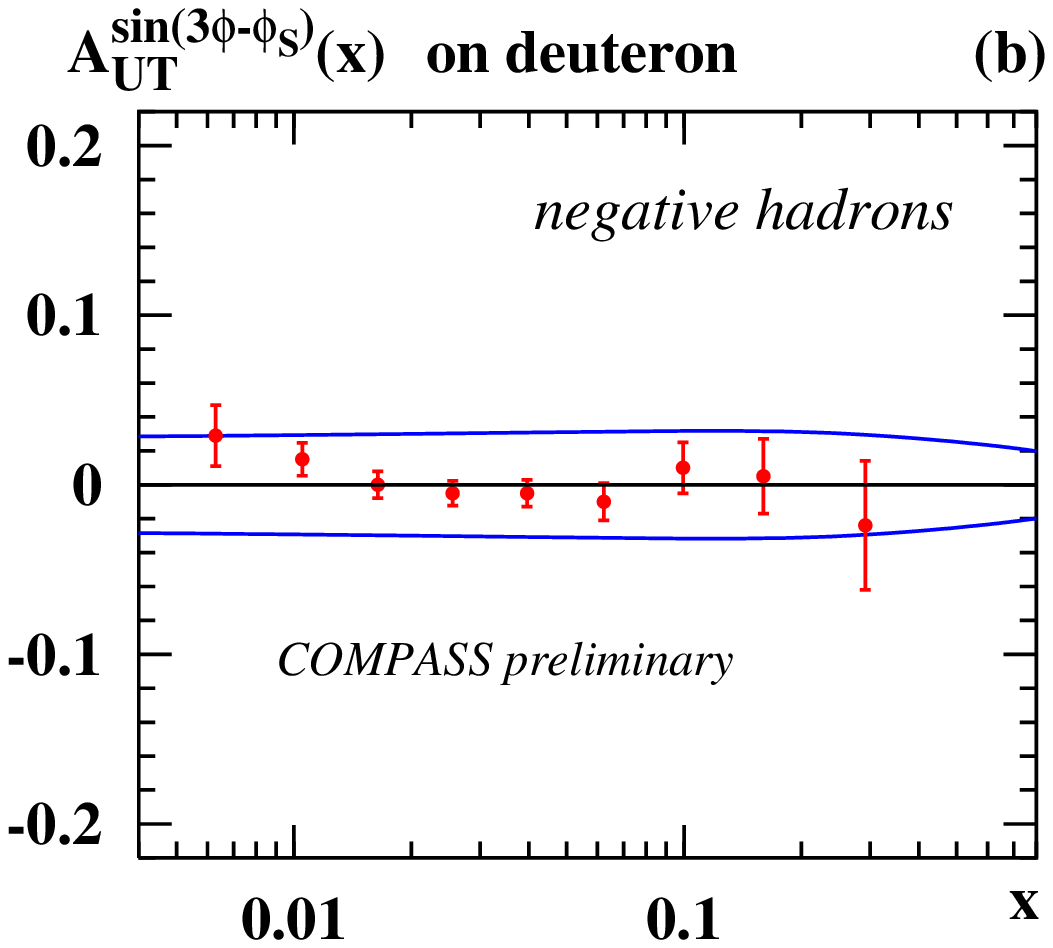}&
\hspace{-8mm}
\raisebox{-1mm}{\includegraphics[width=46mm]
{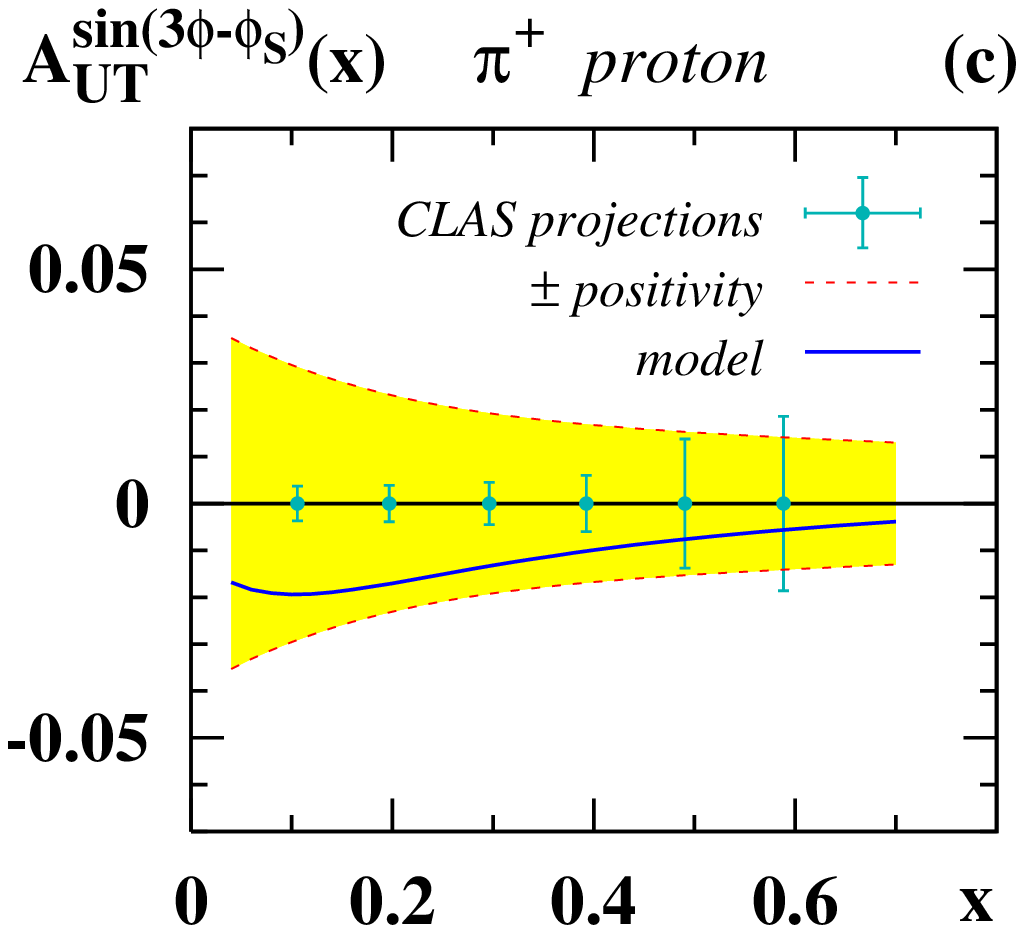}}
\end{tabular}
\caption{\label{Fig:AUT-COMPASS-CLAS-x}
The transverse target SSA $A_{\rm UT,\,\pi}^{\sin(3\phi-\phi_S)}$ for 
deuteron estimated on the basis of the positivity bound vs. 
preliminary COMPASS data \cite{Kotzinian:2007uv} for positive 
{\bf(a)} and for negative {\bf(b)} hadrons.\newline {\bf(c)} The 
same for the $\pi^+$ production from proton in the kinematics of 
CLAS 12 as function of $x$ (error projections from 
\cite{avak-clas12}). Solid curve presents the prediction of 
relativistic covariant model \cite{Efremov:2008mp}. The shaded 
area is the region allowed by positivity (\ref{Eq:positivity}).}
\end{figure}

The results are shown in Fig.~\ref{Fig:AUT-COMPASS-CLAS-x}a,b and 
compared to the preliminary data \cite{Kotzinian:2007uv}. At 
small $x$ the preliminary data favor that pretzelosity does not 
reach the bound. Whether due to the expected suppression at small 
$x$ or opposite signs of $u$ and $d$-flavors, see Sec.~3, cannot 
be concluded.

The important observation is that preliminary COMPASS data 
\cite{Kotzinian:2007uv} do not exclude a sizeable effect in the 
region $x>0.1$, see Fig.~\ref{Fig:AUT-COMPASS-CLAS-x}a,b, where 
JLab can measure with precision. This is demonstrated in 
Fig.~\ref{Fig:AUT-COMPASS-CLAS-x}c showing the $\pi^+$ SSA from a 
proton target in the kinematics of CLAS with 12\,GeV beam upgrade 
(with error projections for 2000 hours run time 
\cite{avak-clas12}). 

It could even more promising to look at SSAs due to Collins 
effect, like $A_{\rm UT}^{\sin(3\phi-\phi_S)}$, in kaon production. 
The statistics for kaon production is lower than for pion 
production, but the SSA might be larger as it is suggested by a 
model \cite{Bacchetta:2007wc} of the Collins function. With a 
RICH detector at CLAS the kaon SSAs could be measured in the 
valence-$x$ region \cite{avak-clas12}.


\section{Conclusions}
\label{Sec-7:conclusions}

The longitudinal SSAs in SIDIS 
\cite{Airapetian:1999tv,Airapetian:2001eg,Airapetian:2002mf,Avakian:2003pk} 
were subject to intensive, early studies 
\cite{DeSanctis:2000fh,Anselmino:2000mb,Efremov:2001cz,Efremov:2001ia,Ma:2002ns} 
that were based on assumptions concerning the flavour dependence 
of $H_1^\perp$ 
\cite{Efremov:2001cz,Bacchetta:2002tk,Efremov:2003eq} that turned 
out not to be supported by data on the Collins effect from SIDIS 
with transverse target polarization 
\cite{Airapetian:2004tw,Alexakhin:2005iw,Diefenthaler:2005gx,Ageev:2006da} 
and $e^+e^-$-annihilations \cite{Abe:2005zx,Ogawa:2006bm}. These 
data  give rise to a new, consistent picture of $H_1^\perp$ 
\cite{Vogelsang:2005cs,Efremov:2006qm,Anselmino:2007fs} which 
invites reanalyses of longitudinal SSAs.

In this work we did this for $A_{\rm UL}^{\sin2\phi} \propto 
\sum_ae_a^2h_{\rm 1L}^{\perp(1)a}H_1^{\perp a}$ from the particular 
point of view of the question whether there are useful, 
approximate relations among different TMDs. In fact, QCD equations 
of motion relate the TMDs entering this SSA to $h_{\rm L}^a(x)$ and 
certain pure twist-3 (and quark mass) terms. Neglecting such 
terms yields an approximation for $h_{\rm 1L}^{\perp(1)a}$ similar in 
spirit to the WW-approximation for $g_{\rm T}^a(x)$ that is supported 
by data.

Our study reveals that data do not exclude the possibility that 
such WW-type approximations work. As a byproduct we observe that 
data on the two SSAs, $A_{\rm UL}^{\sin2\phi}$ and 
$A_{\rm UT}^{\sin(\phi+\phi_S)}$, are compatible. This is important 
because both observables are due to (the same!) Collins effect.

In Ref.~\cite{Kotzinian:2006dw} predictions for 
$A_{\rm LT}^{\cos(\phi-\phi_S)} \propto 
\sum_ae_a^2g_{\rm 1T}^{(1)a}D_1^a$ were made assuming the validity of 
a~WW-type approximation for the relevant pdf. Comparing these 
predictions to preliminary COMPASS data \cite{Kotzinian:2007uv} 
one arrives at the same conclusion. Also here data do not exclude 
the possibility that the WW-type approximation works.

In order to make more definite statements precise measurements of 
these SSAs are necessary, preferably in the region around $x\approx  
0.3 $ where the SSAs are largest. An order of magnitude more data 
on target SSA expected from the CLAS upcoming run \cite{avak-clas6} 
will certainly improve our current understanding of this and 
other SSAs and shed light on spin-orbit correlations.

The value of precise $A_{\rm UL}^{\sin2\phi}$ data should not be 
underestimated. This SSA is in any case an independent source of 
information on the Collins effect. An experimental confirmation 
of the utility of the WW-type approximation 
(\ref{Eq:WW-approx-h1L}), however, would mean that it is possible 
to extract information on transversity, via 
(\ref{Eq:WW-approx-h1L}), from a longitudinally polarized target.


Another subject of this lecture were the properties of the 
pretzelosity distribution function $h_{\rm 1T}^{\perp}$, and the 
presentation of a study of this TMD in the bag model. In the bag and 
in some other quark models we observed an interesting relation, 
which can be summarized for illustrative purposes by the following 
assertion:
\be\label{Eq:relation-illust}
    \mbox{helicity} - \mbox{transversity} = \mbox{
     transverse moment of pretzelosity}.
\ee
That the difference between the helicity and transversity 
distributions is `a measure of relativistic effects' is known 
since long ago \cite{Jaffe:1991ra} (and was also recognized in a 
bag model calculation). However, now we are in a position to make 
this statement more precise. This difference is {just  
$h_{\rm 1T}^{\perp(1)}$ which thus 'measures'} relativistic effects 
in the nucleon, and vanishes in the non-relativistic limit where 
helicity and transversity distributions become equal.

This relation is not supported in models with explicit gluon 
degrees of freedom \cite{Meissner:2007rx}, and, of course, cannot 
be true in QCD where all TMDs are linearly independent. 
Nevertheless, the relation (\ref{Eq:relation-illust}), see 
Eq.~(\ref{Eq:measure-of-relativity}) for its precise formulation, 
could turn out to be a useful approximation. In view of the 
numerous novel functions involved, any well-motivated 
approximation is welcome and valuable \cite{Metz:2008ib}.

Besides being useful for extending our intuition on relativistic 
spin-orbit effects in the nucleon 
\cite{Miller:2007ae,Burkardt:2007rv}, the relation 
(\ref{Eq:relation-illust}) has also an important consequence on 
transversity. {In all quark models where 
(\ref{Eq:relation-illust}) holds}, $h_{\rm 1T}^{\perp u}$ is 
negative. Since $g_1^u(x)$ is positive, this implies that 
$h_1^u(x)>g_1^u(x)$. For the $d$-flavor signs are reversed, but 
in any case $|h_1^a(x)|>|g_1^a(x)|$ which is confirmed in models, 
e.g., \cite{Efremov:2004tz,Schweitzer:2001sr}.

In the bag model, the negative sign of $h_{\rm 1T}^{\perp u}$ arises 
because it is proportional to {\sl minus} the square of the 
$p$-wave component of the nucleon wave function. Thus, in models 
with no higher orbital momentum ($d$-wave, etc.) components, 
$h_{\rm 1T}^{\perp u}$ is manifestly negative ($h_{\rm 1T}^{\perp d}$ has 
opposite sign dictated by SU(6) symmetry, and predicted in large 
$N_c$ \cite{Pobylitsa:2003ty}).

This prediction can be tested at JLab. Since the production of 
positive pions from a proton target is dominated by the 
$u$-flavor, one expects a negative $\sin(3\phi-\phi_S)$ SSA, see 
Fig.~\ref{Fig:AUT-COMPASS-CLAS-x}.

Forthcoming analyzes and experiments at COMPASS, HERMES and JLab 
\cite{avak-clas6,avak-clas12} will provide valuable information 
on the pretzelosity distribution function, and will deepen our 
understanding of the nucleon structure.

 \section*{Acknowledgments }
The work is supported by BMBF, German--Russian collaboration 
(DFG-RFFI, 436 RUS 113/  881/0), the EIIIHP project RII3- 
CT-2004-506078, the Grants RFBR 09-02-01149 and 07-02-91557, RF 
MSE RNP.2.2.2.2.6546 (MIREA), the Heisenberg--Landau Program of 
JINR, the contract DE-AC05-06OR\-23177.


\end{document}